# Multi-Scale Experimental Characterization for LS-DYNA MAT213 Modeling of Composite Structures under High Strain Rate

Jackob Black[1], Ryan Premo[2], Robert K. Goldberg, F. ASCE[3],
Trenton M. Ricks[4], Troy Lyons[5], and Han-Gyu Kim[6]

[1]Mississippi State University, Department of Aerospace Engineering, Mississippi State, MS 39762; email: jwb774@msstate.edu
[2]Mississippi State University, Department of Aerospace Engineering, Mississippi State, MS 39762; email: rgp103@msstate.edu
[3]NASA Glenn Research Center, Ceramic and Polymer Composites Branch, Cleveland, OH 44135; email: robert.goldberg@nasa.gov
[4]NASA Glenn Research Center, Multiscale and Multiphysics Modeling Branch, Cleveland, OH 44135; email: trenton.m.ricks@nasa.gov
[5]NASA Glenn Research Center, Structural Dynamics Branch, Cleveland, OH 44135; email: troy.lyons@nasa.gov
[6]Mississippi State University, Department of Aerospace Engineering, Mississippi State, MS 39762; email: hkim.ae@msstate.edu

**ABSTRACT**

Aerospace structures often experience high strain rate events such as ballistic impact, crash, or crush. A material model has been developed that enhances the capability to simulate the dynamic response of composite materials under these loading conditions. The material model has been implemented into the commercially available transient dynamic finite element code LS-DYNA as MAT213. The model can simulate the nonlinear deformation, damage, and failure that takes place in a composite under dynamic loading conditions. The specific goal of this work is to characterize the MAT213 input for the representative material. The specific composite material being examined consists of T700G unidirectional carbon fibers and a low-melt PolyArylEtherKetone (LMPAEK) thermoplastic resin system. It is formally referred to as Toray TC1225 LMPAEK T700G. As the initial part of this work, this paper is focused on characterizing the material parameters for the MAT213 deformation model based on results obtained from multi-scale experimentation. The effort concentrated on characterizing the in-plane material response suitable for use with thin shell elements. For shell elements within MAT213, tabulated stress-strain results from tension and compression tests in the longitudinal and transverse directions and in-plane shear tests are required. Due to the difficulty of measuring small strains in the transverse direction, a multi-scale testing method was developed. Macro-scale testing is performed per the typical ASTM methods while micro-scale testing uses a microscope along with smaller coupon sizes to obtain the smaller strains in the transverse direction of each test. For both testing methods, a VIC-2D camera and software for DIC (Digital Image Correlation) analysis are used. Using the DIC combined with each test fixture, reliable stress and strain data are collected. This material was characterized based on this data. Single-element simulations are then executed to validate that the results from the simulation match the input curves. Following this process, coupon-level models for the test sets are created for analysis using MAT213. The results from the analytical simulations are compared to the experimental results obtained in this study to validate the model. Ultimately this effort aims to improve the ability to simulate the dynamic response of composite materials.

# INTRODUCTION

As composite materials are gaining increased use in the aircraft industry where impact resistance under high-energy impact conditions is important, the demand for reliable and accurate material models to simulate the damage, deformation, and failure response of polymer matrix composites under impact conditions is becoming imperative. LS-DYNA is a commercially available transient dynamic finite-element code extensively used in the aerospace industry. There are several material models within LS-DYNA currently available for application to the analysis of composites. One of the material models in this finite element code is called MAT213.

To develop MAT213, a multi-institution consortium was formed to develop and implement a novel, three-dimensional elastoplastic with damage composite material model within LS-DYNA In many of the existing composite models in LS-DYNA, only pointwise data such as modulus and failure stress are input into the model. This type of approach results in a crude approximation, at best, to the actual stress-strain curve. An improved approach, employed in MAT213, uses tabulated data from a well-defined set of experiments to define the complete stress-strain response of the material (Goldberg, 2018). MAT 213 can be used with thin shell, membrane, or solid elements. For cases where all of the required experimental data is not available, numerical methods (which could potentially include machine learning algorithms in the future) can be used to fill in the needed data. While not discussed in this paper, a damage model is implemented into MAT 213 which accounts for the nonlinear unloading and post-peak stress degradation observed in composites without regard to the specific constituent level damage mechanisms. For this study, the Toray TC1225 LMPAEK composite (composed of T700G unidirectional carbon fibers and a low-melt PolyArylEtherKetone (LMPAEK) thermoplastic resin system) was chosen. The Toray TC1225 LMPAEK T700G has a high strength and improved performance under extreme thermal conditions. The LMPAEK material was already tested and characterized by the National Center for Advanced Materials Performance (NCAMP) (Clarkson, 2020). However, an enhanced set of data, including tabulated stress-strain curves, is required for appropriate input into MAT213. The first task described in this paper was to gather appropriate data from experimental testing required for input into MAT213. Once the experimental data was gathered, a MAT213 material card was generated using procedures to be described here. The developed MAT213 material card was then used to conduct single-element simulations and coupon-level simulations to validate the MAT213 input. This study is only concentrating on the baseline quasi-static tests and analyses required to characterize the deformation model for the studied material. Characterization of the MAT 213 damage and failure models and simulations of impact, crash, and crush problems will be conducted in future work.

# EXPERIMENTAL SETUP AND METHODOLOGY

For shell elements within MAT213, tabulated stress-strain results from tension and compression tests in the longitudinal (1) and transverse (2) directions and in-plane (1-2 direction) shear tests are required. When using DIC, it can be difficult to measure small strains in the transverse direction, so a multi-scale testing method was developed. The micro-scale testing is performed to acquire high-resolution strain data because it is challenging to obtain high-fidelity data in the transverse direction for some tests. For both testing methods, a Basler machine vision camera and Correlated Solutions VIC-2D software packages for Digital Image Correlation (DIC) analysis are used.

The testing methodology utilizes both the Shimadzu AGX-V test frame with the accompanying load cell, as well as the Psylotech µTS microscopic test setup, as can be seen in Figure 1. Both testing environments, macro- and microscopic, utilize the same Basler Ace 12MP machine vision camera, differing only by changing lenses between test environments. ASTM methods pertaining to the performed tests are listed in Table 1 below.

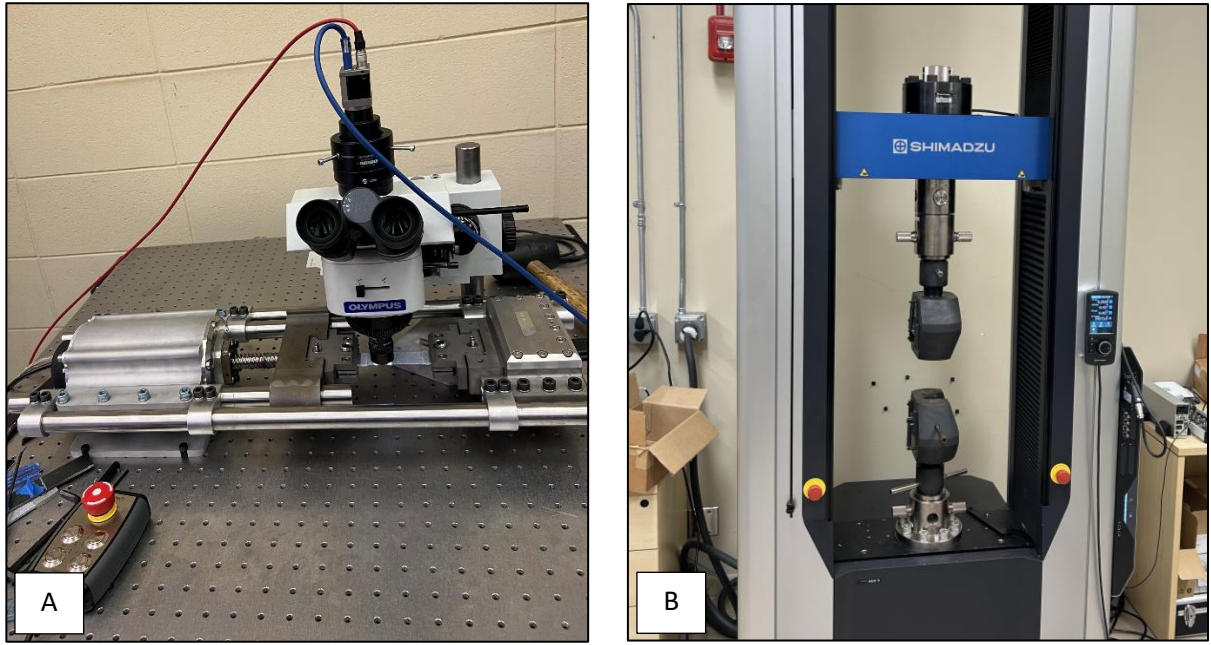

*Figure 1: (a) The Psylotech µTS microscopic test setup and (b) The Shimadzu AGX-V test frame with accompanying load cell.*

*Table 1. ASTM test methods*

| Test Type | ASTM Method |
|---|---|
| 0° and 90° Tension | D3039/D3039M-17 (ASTM International, 2017) |
| 0° and 90° Compression | D6641/D6641M-16e2 (ASTM International, 2021) |
| 0° V-Notched Shear | D5379/D5379M-19e1 (ASTM International, 2021) |

**EXPERIMENTAL RESULTS**

The data acquired through both macro and micro-level testing, seen in Figure 2. below, closely mirrors published data from NCAMP. To compare the experimental data gathered to the NCAMP data, the modulus and strength values from each test were compared. The maximum deviation from the median published value is 11%, and all recorded material properties are within the standard deviation also published by NCAMP (Clarkson, 2020). This comparison was done to prove that the experiments were performed properly and to verify that the gathered data is fit for use in the simulations.

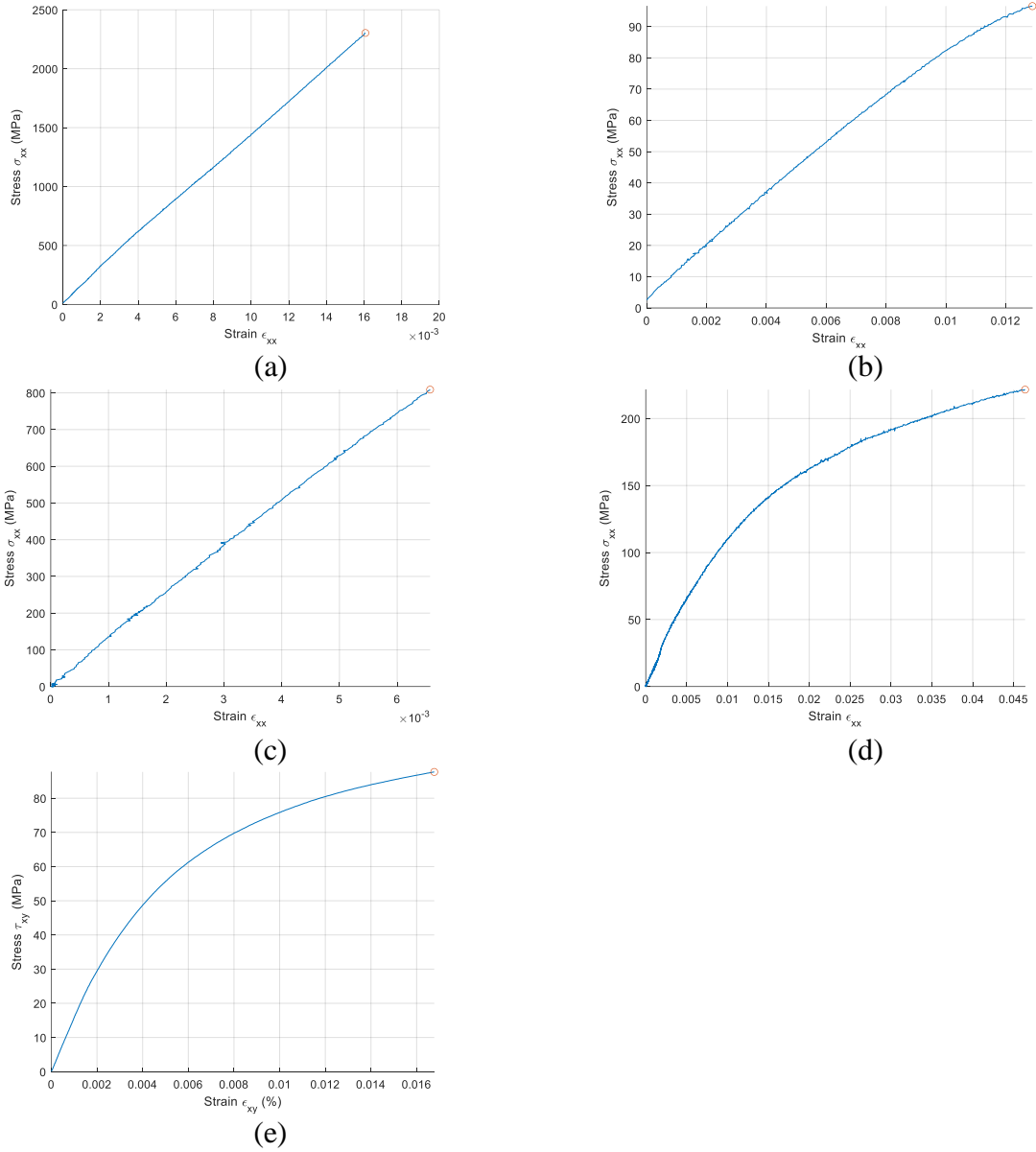

*Figure 2: Experimental stress-strain data (a) Axial (longitudinal) stress-strain 0° Tension (b) Axial (transverse) stress-strain 90° Tension (c) Axial (longitudinal) stress-strain 0° compression (d) Axial (transverse) stress-strain 90° compression (e) Shear strain 0° V-notch shear*

A set of elastic properties is required for each material input into LS-DYNA. These properties are gathered from the experiments mentioned previously, and the results can be seen below in Table 2.

*Table 2. Material properties*

| | |
|---|---|
| Material Density (g/cm$^3$) | 0.001579 |
| 0° direction Modulus (N/m$^2$) | 144700.0 |
| 90° direction modulus (N/m$^2$) | 5581.42561 |
| Poisson's Ratio (Unitless) | 0.07084380 |
| Shear Modulus (N/m$^2$) | 7416.07 |

**MODELING**

MAT002 is a linear orthotropic material model in LS-DYNA, and it is considerably simpler than MAT213. Single-element simulations were first run with the MAT002 model to ensure that the basic elastic response of the material was being simulated correctly. These preliminary single-element simulations motivated both damping and hourglass control to be invoked for numerical stability. Some changes to the boundary conditions (switching from displacement to traction control) in the shear element were also made to create the desired stress state.

**SINGLE-ELEMENT MODELING USING MAT002**

Each element in the single-element simulations is loaded to create a desired stress state. As shown in Figure 3, the red and blue elements on the top left undergo 1 direction tension and 1 direction compression, respectively. The dark green and yellow elements on the upper right undergo 2-direction tension and 2-direction compression, respectively. The lighter green element on the bottom right is put into a pure shear stress state in the 1-2 plane. All of the elements on the top have their bottom nodes constrained from vertical displacement, and their nodes on the top are driven vertically. Horizontal motion is unrestricted to maintain a uniaxial stress state. The element undergoing shear on the bottom right has its material axes rotated 45 degrees from the horizontal. Forces to create horizontal compression and vertical tension are applied. In the material. coordinate system, the stress state created by those forces result in an approximately pure-shear stress state. To ensure the validity of this model the stress-strain curves from each condition were plotted for each of the elements. It was then determined whether each element displayed the appropriate elastic properties (such as modulus) compared to the input data. Once this was completed the MAT213 model creation was started. To ensure the conciseness of this paper, the results from each single-element simulation are not included.

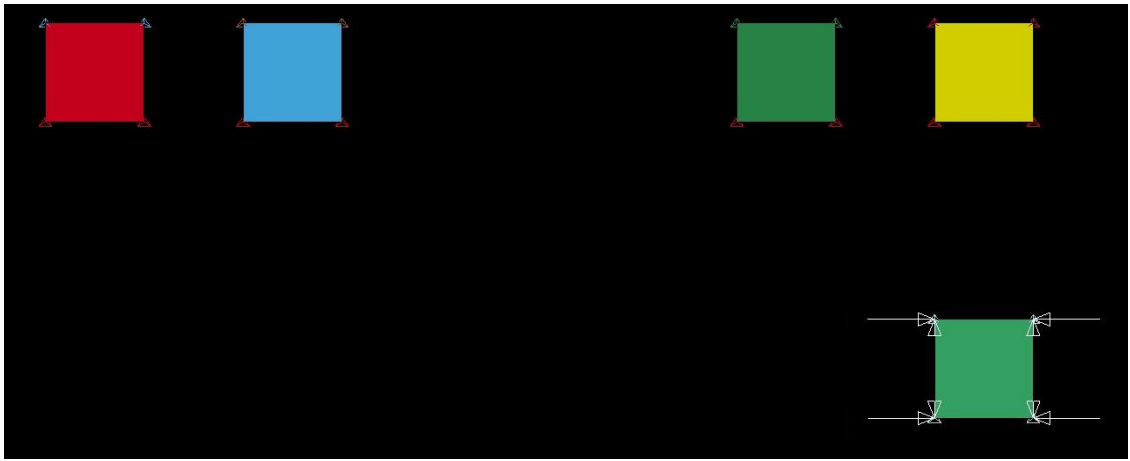

*Figure 3. Single-element simulation showing loading and boundary conditions*

**SINGLE-ELEMENT MODELING USING MAT213**

Following the MAT002 single-element simulations, an additional series of single-element simulations were conducted using MAT213 as the material model. The stress-strain curves obtained from the single-element simulations were then compared to the input stress-strain curves to ensure that the material model with the specified parameters was performing as intended. The authors have found that the intermediate steps of going from MAT002 single-element simulations, to MAT213 single-element simulations, and then afterwards to coupon-level simulations ultimately save the user time.

MAT213 uses tabulated stress-strain data obtained from the experimental tests described earlier as input. However, the experimental stress-strain data used in MAT213 had to be cleaned prior to use in the MAT213 input deck to remove experimental noise from the data. The data was made smoother, ensured to be monotonically increasing, and adjusted as needed to have a slope in the plastic region that is less than the slope in the elastic region. In all cases, the adjusted curves were set to fit as closely to the actual experimental stress-strain data as possible.

**COUPON-LEVEL MODELING USING MAT213**

Once the single-element simulations were completed, a series of coupon-level simulations were conducted to ensure that the specified input parameters such as the plasticity flow rule coefficients, which can be seen below in Table 3, and the curves for the MAT213 input deck accurately represented the material response in a multi-element model. Each coupon is modeled with Hughes-Liu 4-node shell elements in LS-DYNA. Simulations of the 0- and 90-degree tests were conducted. Similar to the single-element simulations, each of the tension and compression elements was fixed on the bottom and the displacement was applied to the top as shown in Figure 4. The simulations concentrated on modeling the straight-sided gage section of the tests in order to ensure only uniaxial stresses were present. In addition, by limiting the model in this fashion a relatively coarse mesh was able to be used which lowered the computational cost of the simulations. The shear coupon simulation has not yet been performed and will be conducted in the future.

*Table 3. Flow rule coefficients*

| H11 | 0.1 |
|---|---|
| H22 | 1.0 |
| H44 | 5.0 |

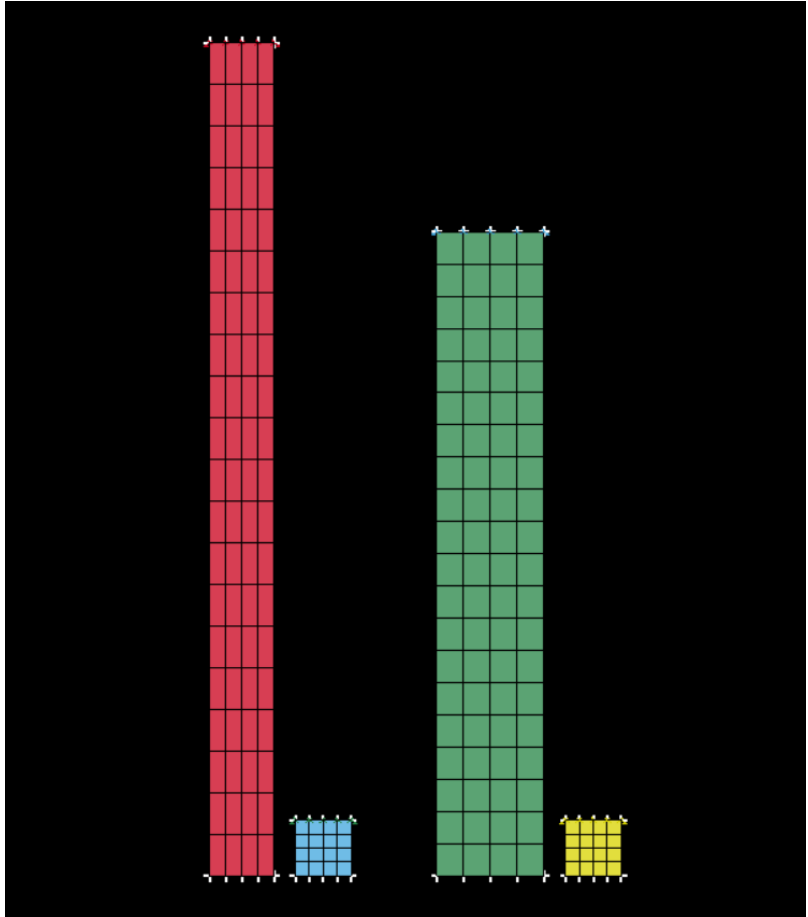

*Figure 4: Coupon simulation featuring zero-degree and ninety-degree tension and compression tests*

## SIMULATION RESULTS

The coupon-level experiments described earlier in this paper were simulated using the MAT213 material model and the numerical results are shown below in Figures 5-8. The zero-degree tension and compression simulations are seen in Figures 5 and 6. The stress-strain curves from the MAT213 simulation, depicted as a blue line, are consistent with what is seen in the experimental data, depicted as a red dashed line. Figures 7 and 8 depict the ninety-degree tension and compression tests. The results shown in Figure 5 show some variance between the experimental data and the simulation results. This is because the curves generated for MAT213 input are required to have a slope in the plastic region that is less than the slope in the elastic region. The simulations were able to simulate the material response comparable to the experimental data for the conditions examined. Therefore, this input card is acceptable for use in further applications.

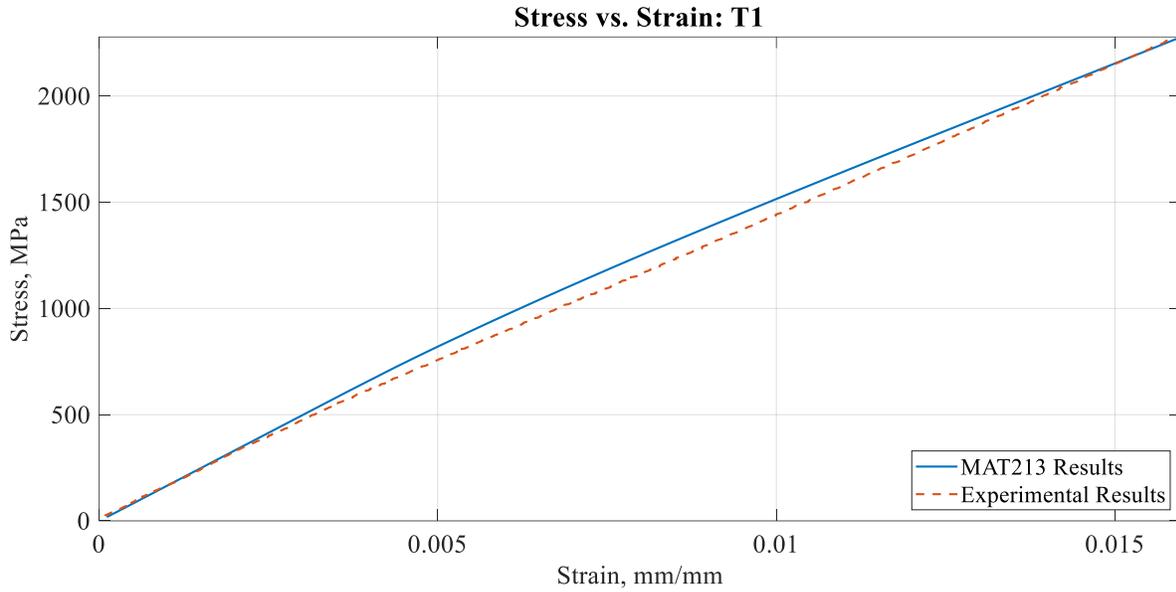

*Figure 5: Stress-strain curves for the zero-degree tension test where the red dashed line is the experimental curve and the solid blue line is the results from the MAT213 simulation*

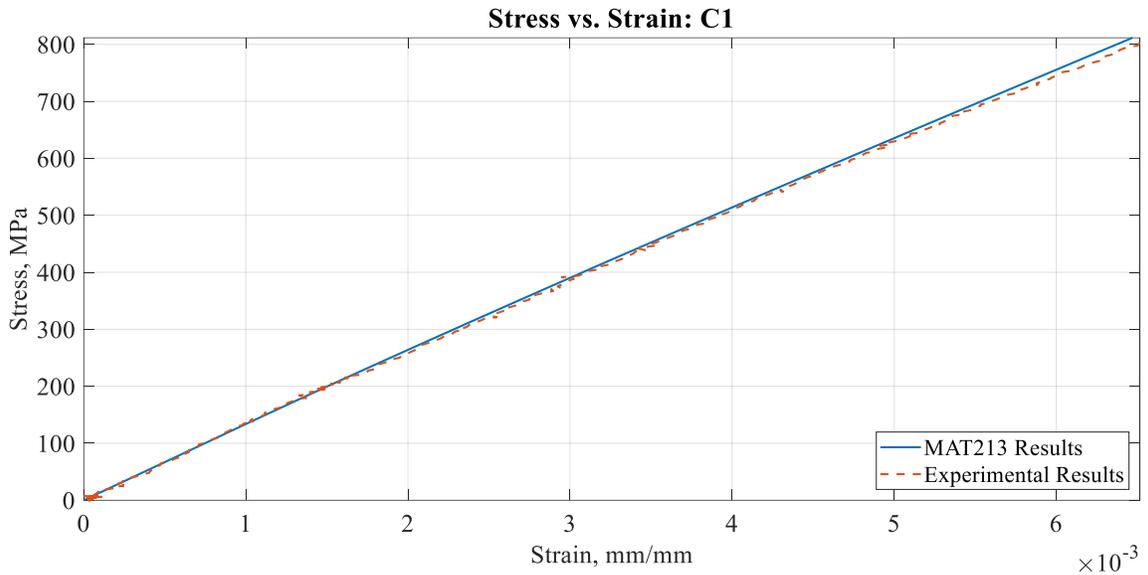

*Figure 6 Stress-strain curves for the zero-degree compression test where the red dashed line is the experimental curve and the solid blue line is the results from the MAT213 simulation*

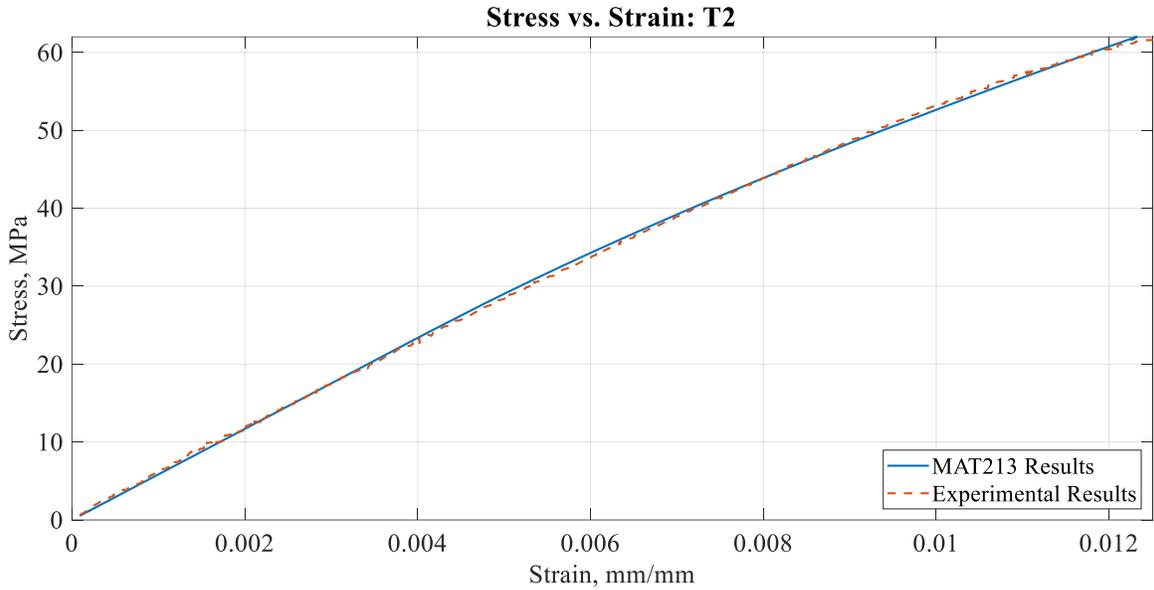

*Figure 7: Stress-strain curves for the ninety-degree tension test where the red dashed line is the experimental curve and the solid blue line is the results from the MAT213 simulation*

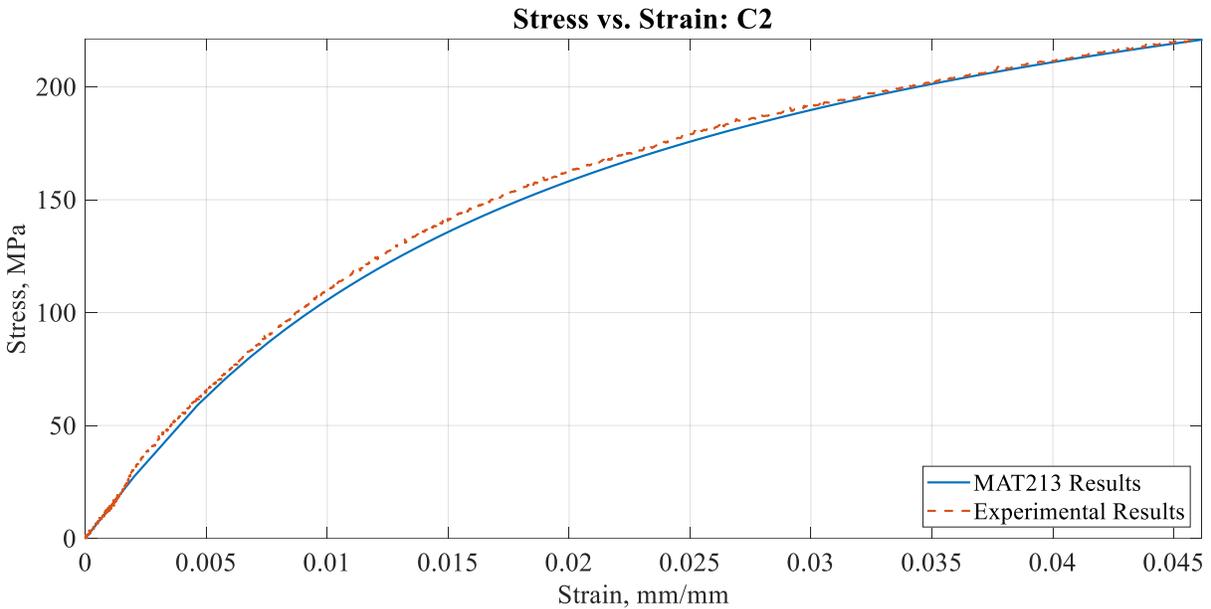

*Figure 8: Stress-strain curves for the ninety-degree compression test where the red dashed line is the experimental curve and the solid blue line is the results from the MAT213 simulation*

**CONCLUSION/FUTURE WORK**

This report described the process of gathering experimental stress-strain curves and using that data to create and verify an input deck suitable for use with the LS-DYNA MAT213 material model for a composite material consisting of T700G unidirectional carbon fibers and a low-melt

PolyArylEtherKetone (LMPAEK) thermoplastic resin system called Toray TC1225 LMPAEK T700G. The elastic properties obtained from the material tests closely matched data gathered by the National Center for Advanced Materials Performance (NCAMP). The tabulated data gathered from the experimental tests was used to develop working single-element models using the MAT002 and MAT213 material models, as well as coupon simulations using MAT213 which generated results similar to the experimental data gathered. This work demonstrates the ability of MAT 213 to appropriately simulate the deformation response of a thermoplastic matrix composite. Future work in this effort will build towards augmenting the developed input deck to allow for the simulation of ballistic impact damages of a structure composed of this material. The quasi-static simulations will provide the groundwork for the implementation of the continuum-level damage sub-model in MAT213. Rate and temperature-dependent impact tests will be performed to gather the necessary data to simulate the dynamic response of structures seen in more realistic cases.

## ACKNOWLEDGMENTS

The authors gratefully acknowledge the support of NASA (Grant No.: 80NSSC22M0206) as well as Sandi Miller and Mike Pereira of the NASA Glenn Research Center.